\documentclass[aps,preprint,showpacs]{revtex4}
\begin{document}  %
\draft
\input{psfig.tex}
\title{Possible Structures of Sprites } 
\author{Kwang-Hua W. Chu}  
\affiliation{P.O. Pox 39, Tou-Di-Ban, Road XiHong, Wulumuqi 830000, PR China  }
\begin{abstract}
Upon using the hydrodynamic analog we can derive
some families of stationary  Beltrami field-like solutions
from the free
Maxwell equations in vacuum. These stationary
electromagnetic fields are  helical and/or column-like
once they are represented in a suitable frame of reference.
Possible dendritic and jelly-fish-like patterns of sprites are demonstrated.
%
%
%
\end{abstract}             %
\maketitle
\bibliographystyle{plain}

\noindent           
The majority of the great discoveries in physics have
been made by chance (electromagnetism, radioactivity, etc.) and only
in some rare cases was the experiment preceded by theory (lasers,
thermonuclear reactors). It seems likely that the problem of sprites
 belongs to the latter one.
Since their discovery by Franz {\it et. al.} [1], sprites have
now been observed over thunderstorms all over the world [2-4].
A spectacular manifestation of these sprites :
the intense transient quasistatic
electric (QE) fields of up to $\sim 1$ kV $\cdot \,m^{-1}$, which
for positive (Cloud-to-Ground) CG discharges is directed downwards,
can avalanche accelerate upward-driven
runaway MeV electron beams, producing brief ($\sim 1$ ms) flashes of
gamma radiation.
These intense fields are large luminous
discharges in the altitude range of $\sim$ 40 km to 90 km,
which are produced by the
heating
of ambient electrons for a few to tens of milliseconds following
intense lightning
flashes. \newline
Thousands of positive ground flash-sprite
associations have been identified through comparisons with
video imaging/optical sensor verification of sprites.
Beginning with suggestions by Wilson [5], the
electrostatic field change of the lightning flashwas sufficient to
exceed the dielectric strength of the mesosphere and initiate the
sprite. This Wilson mechanism for sprites initiated by
conventional dielectric breakdown is polarity independent-positive and
negative changes in charge moment change in
excess of the threshold should be equally effective in the initiation of sprites.
Sprites  require a lowly or nonionized
medium; they can propagate from the ionosphere
downwards or from some lower base upwards or they can
emerge at some immediate height and propagate upwards and
downwards [6-7]. %
\newline
The first columniform sprites (c-sprites) reported [8] were vertical
columns of light that extended from
about 76 to 87 km and probably less than 1 km in diameter.
According to Wescott {\it et al.} [8] the uncertainty in height determination
for their c-sprite measurements was approximately
1 pixel or less than 1 km. The standard deviation of the heights
for the 28 triangulated images was 1.9 km for the top and 1.4 km
for the bottom.
Additional observations of c-sprites, some of
which may extend to lower altitudes, have been recently reported
[9-10]. The bodies of
c-sprites usually appear as nearly straight lines, in some cases
bent slightly, and are composed of bright beads and dark regions.
 \newline
Meanwhile, some researchers [7-8] described
the phenomenon, called
{\it red Sprites}, as a luminous column that stretches
between 50 and 90 km, with
peak luminosity in the vicinity of 70-80 km. The flashes have an average
lifetime
of a few milliseconds and an optical intensity of about 100 kR. Red Sprites
are associated with the presence of massive thunderstorm clouds, and among the
most puzzling aspects of the observations is the presence of spatial structure in
the emissions reported by Winckler et al. [7].
Vertical striations, denominated
'tendrils', with horizontal size of 1 km or smaller,
often limited by the instrumental
resolution, are apparent in the red Sprite emissions [9-10]. \newline
As one example, the key features of the previous model [9] are: (1) an ambient
conductivity profile that falls between a measured nighttime and a
measured daytime conductivity; (2) an aerosol reduced conductivity
in a trail from a meteor that passed through some time during the
evening, and (3) a cloud-to-ground (hereafter CG) lightning stroke,
with sufficient charge transfer, subsequent to and occurring within
an hour of the development of the reduced conductivity trail. This
model predicts a temporally brief column of light resulting from the
conventional breakdown of air in a strong electric field in the
observed altitude range. The results  we shall describe in this short paper might provide an
explanation for the observed c-sprite phenomena.\newline
The free Maxwell equations
(FME)  are
\begin{displaymath}
 \epsilon_{ijk} \partial_j E_k =-\frac{1}{c}\frac{\partial B_i}{\partial t},
 \hspace*{12mm} \epsilon_{ijk} \partial_j B_k =\frac{1}{c}\frac{\partial E_i}{\partial t},
\end{displaymath}
\begin{displaymath}
 \partial_i E_i =0,\hspace*{12mm} \partial_i B_i =0,
\end{displaymath}
where $\partial_i \equiv \partial/\partial x_i$, $x_i=(x,y,z)$,
$\epsilon_{ijk}$ is an antisymmetric tensor or symbol.
We presume that the solutions of FME are
\begin{equation}
 E({\bf r},t)={\bf e}({\bf r}) {\cal F}(t), \hspace*{12mm} \mbox{and}
 \hspace*{12mm} B({\bf r},t)={\bf b}({\bf r}) {\cal G}(t).
\end{equation}
In order to obtain solutions of this system with three constants
only and to obtain sinusoidal solutions, we also presume that
\begin{equation}
 \frac{-d {\cal G}(t)/dt}{{\cal F}(t)}=\frac{d {\cal F}(t)/dt}
 {{\cal G}(t)}=\Omega.
\end{equation}
Therefore, the general solutions of above system [11] are
\begin{equation}
 {\cal F}(t)=C_0 \sin (\Omega t-\phi), \hspace*{12mm} \mbox{and}
  \hspace*{12mm} {\cal G}(t)=C_0 \cos (\Omega
 t-\phi),
\end{equation}
where $C_0$ and $\phi$ are arbitrary constants. Now, equations for
${\bf e}$ and ${\bf b}$ become
\begin{equation}
 \nabla \times {\bf e}=\frac{\Omega}{c} {\bf b}, \hspace*{24mm}
 \nabla \times {\bf b}=\frac{\Omega}{c} {\bf e}.
\end{equation}
Above equations give
\begin{equation}
 \nabla \times {\cal A} = \frac{\Omega}{c} \,{\cal A}, \hspace*{24mm}
\mbox{with} \hspace*{6mm} {\cal A}={\bf e}+{\bf b},
\end{equation}
where  ${\bf e}$ is a polar vector (linked to the electric field
${\bf E}$ in space, ${\bf E}={\bf e}\, {\cal F}(t)$), ${\bf b}$ is an
axial vector (linked to the magnetic field ${\bf B}$ in space,
${\bf B}={\bf b}\, {\cal G}(t)$).
\newline Now, from the vector analysis and the hydrodynamic analogy
(e.g., for a velocity field ${\bf u}$, we have the vorticity field
${\mathbf \omega}$  which is defined as $\nabla \times {\bf u}$), we
know that
\begin{equation}
 (\nabla \times {\cal A}) \parallel {\cal A} \hspace*{6mm} \mbox{or}
 \hspace*{6mm} (\nabla \times {\cal A}) \times {\cal A}={\bf 0},
\end{equation}
here, $\parallel$ means 'parallel to'. In fact, people termed
${\bf L}=(\nabla \times {\cal A}) \times {\cal A}$ as a {\it Lamb}
vector [12-13].\newline
In generalized fluid mechanics, we call
the flow which satisfies
\begin{equation}
 {\bf \omega} \times {\bf u} = {\bf 0}
\end{equation}
a {\it Beltrami} flow [11-14], sometimes a {\it screw} motion  or
helical flow  [11-13] (here, ${\bf \omega} \not = {\bf 0}$).
\newline With regard to these, vector fields satisfying Eq.
(7) are helical or screw fields. It means ${\cal A}$ or the
combination of ${\bf e}$ and ${\bf b}$ is a helical or screw field.
One class of these solutions can be obtained and expressed in
cylindrical polar coordinates $(r,\theta,z)$ as
\begin{equation}
 u=\bar{A}_0 r \cos(\zeta),  \hspace*{6mm} v=\hat{\Omega} r+ \bar{A}_0 r \sin
 (\zeta), \hspace*{6mm} w=-\frac{2 \bar{A}_0}{k} \sin(\zeta),
\end{equation}
where $\zeta =k(z-c_0 t)=kz-2\hat{\Omega}t$. When ${\bf
u}=(u,\hat{v},w)\equiv {\cal A}$ is expressed in a frame of
reference rotating with the mean angular velocity $\hat{\Omega}$
($\hat{v}= \bar{A}_0 r \sin
 (\zeta)$), we have $\nabla
\times {\bf u}\equiv \omega= -k {\bf u}$ or ${\bf \omega} \times
{\bf u} = {\bf 0}$.  The stationary form of this solution could be
similar to ${\bf e}+{\bf b}$ (i.e., ${\cal A}$). We remind the
readers that eqns. (8) describe a field that grows proportionally to
the distance from the cylindrical axis. Therefore, in unbounded
vacuum, as there is no dissipation, $\bar{A}_0 r \sim O(1)$ as
$r\rightarrow \infty$ should be presumed otherwise they will give
some non-localized electromagnetic field configurations with
infinite energy. Under this consideration, once $r\rightarrow
\infty$, $w$ is rather small (as $\bar{A}_0 \rightarrow 0$ if $k$ is
finite) and this (axial) component (related to magnetic field; will
be illustrated below) behaves like a weak 'thread'! On the other
hand, since $\bar{A}_0 r \sim O(1)$, once $r\rightarrow 0$, there
might be a singularity at $r=0$ ($\bar{A}_0$ has been restricted for
very larger $r$). This singularity resembles that of a free vortex
(vortex core or the filament of a vortex tube). Note that, once the amplitude $\bar{A}_0$ being replaced by
$A(t)$ (with ${d A}/{dt}= -\nu A k^2$, $\nu$
plays the role of damping,
the solution of above equation easily reads $ A(t)=A(0) e^{-\nu
k^2 t}$), we then have
 $u=A(t) r \cos(\zeta)$,  $ \hat{v}= A(t) r \sin
 (\zeta)$, $ w=-{2 A(t)} \sin(\zeta)/{k}$.
To demonstrate
our solutions, we plot those stationary ones of them into following
figures. We shall firstly present the electromagnetic field in
vacuum. Parameters are selected as $\bar{A}_0=0.45, k=2.5$; the maximum
spatial range : $0\le r\le 1$, $0\le z \le 2\pi$. The vector field
is represented in a  rotating  frame of
reference (Fig. 1) with the mean angular velocity $\hat{\Omega}=5.0$.
We can observe the helical and the dendritic  features of our
solutions and the possible pattern of a sprite-like electromagnetic
field.
\newline  The outer and inner isosurface plots (for
the total modulus of ${\cal A}$ and the modulus of $u {\bf i}_r +v
{\bf i}_{\theta}$, respectively) are presented in Fig. 1.
To make sure there are also
possible sprite-like fields in our results as there are so many
parameters needed for tuning and the 3D vector view is sometimes not
comprehensive, thus we only show those isosurface plots below. \newline
With intensive searching, we can find out, at least, for
$\hat{\Omega}=5$, $\bar{A}_0=0.45, k=1.5$, there exists also
possible column-like together with helical field for sprites. This
pattern is shown in Fig. 2  for the same frame of reference (
 in a rotating frame of reference). The outer (green) surface is for
the total modulus of ${\cal A}$ which is already in a helical
shape. The inner (yellow) surface is for the modulus of $u {\bf i}_r +v {\bf
i}_{\theta}$ which demonstrates the possible column-like channel.
As parameters are tuned into
$\hat{\Omega}=0.2$, $\bar{A}_0=1.5, k=2.5$, the dendritic  features
of sprites are illustrated clearly in Fig. 3.
We can now easily capture the characteristics of helical or possible
helix-like fields in Figs. 1, 2 and 3. These results
confirm our present approach (cf. images or figures in [8-9]).
\newline Note that, once we introduce
\begin{equation}
 {\bf e} ({\bf r})=\frac{1}{2}[{\cal A} ({\bf r})-{\cal A} (-{\bf
 r})],  \hspace*{12mm}
{\bf b} ({\bf r})=\frac{1}{2}[{\cal A} ({\bf r})+{\cal A} (-{\bf
 r})],
\end{equation}
considering the definition of Poynting vector ${\bf S}=c({\bf
E}\times {\bf B})/(4 \pi)$, we also can obtain the relevant results
about  ${\bf S}$ (in fact, ${\bf b} ({\bf
r}) \propto w$ as evidenced from above equation). The calculation is
now only related to $ {\cal A}({\bf r})\times {\cal A} (-{\bf r})$
instead of ${\bf e} \times {\bf b}$. Meanwhile, as the energy
density $W$ is defined as $(E^2+B^2)/8 \pi$, we can directly
interpret $W$ from the magnitude or modulus of ${\cal A}$ from Eq.
(8). Meanwhile, from Eqs. (8) and (9), we can obtain the explicit
expression of the electric (${\bf e} ({\bf r})$) and magnetic (${\bf
b} ({\bf r})$)fields.   We might have, the electromagnetic energy
($|{\cal E}| \propto \alpha \gamma R_0^{-3}$, $\alpha=- (\Omega R_0/
c) \cos(\Omega R_0/c)+\sin(\Omega R_0/c)$ and $\gamma=\alpha-
(\Omega^2 R_0^2/c^2) \cos(\Omega R_0/c)$) within spheres of radius
$R_0$, which is the solution of
    $\tan({\Omega R_0}/{c})= {\Omega R_0}/{c}$,
and it is independent on the time.
\newline
However, once  we selected parameters as
$\hat{\Omega}=5$, $\bar{A}_0=1.2, k=0.8$, a single sprite
appears (cf. Fig. 4 in [8] or Fig. 4a in [9] or Figs. 5 and 6 in [15])
which is shown in Fig. 4.
The other interesting case is for $\hat{\Omega}=0.6$, $\bar{A}_0=0.6, k=0.15$
as shown in Fig. 5 : a  jelly-fish-like pattern
with the downward tendrils-like
feature  appears.\newline
To be specific, we like to make a final
remark about the characteristic of our Beltrami fields : Eq. (8)
especially when $k$ is rather small (solutions for $k=0$ violate the
Beltrami condition!). Once $k \rightarrow 0$, considering the limit
of the $w$-component in the stationary case, $w \propto z$ with $z$
being finite. Under this situation, our solutions resemble those
results for high-electric-field cases [15].
\newline
In a brief
summary, the author likes to point out, the helical and column-like
fields illustrated above  are closely linked to the formation of
sprites, however, there will be other unknown, say, relativistic,
effects during the final stage in the formation of  lightning.
We shall consider other more
complicated problems [16-21] in our future works. 

\newpage

\oddsidemargin=-5mm

\pagestyle{myheadings}

\topmargin=-18mm \textwidth=17cm \textheight=27cm


%
%
\vspace{3mm}
\psfig{file=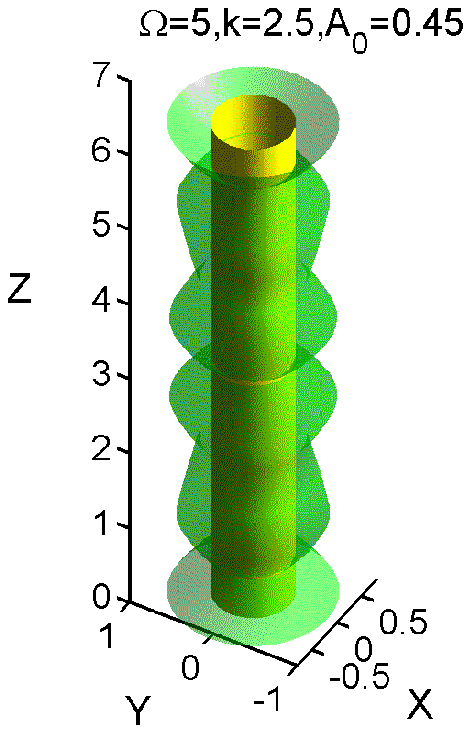,bbllx=0cm,bblly=13.5cm,bburx=15cm,bbury=24cm,rheight=11cm,rwidth=11cm,clip=}
\vspace{3mm}
\begin{figure}[h]
\hspace*{6mm} Fig. 1 \hspace*{1mm} Isosurface plots of the solution
$({\bf e}+{\bf b})$  in a rotating  frame of reference.\newline
\hspace*{7mm}   Possible helical and column-shape electromagnetic
fields. $\hat{\Omega}=5, \bar{A}_0=0.45$, $k=2.5$. \newline
\hspace*{7mm} (the outer or green surface is for $|{\cal A}|$ and
the inner or yellow surface is for $|u {\bf i}_r +v {\bf
i}_{\theta}|$) \newline \hspace*{7mm}
This result resembles qualitatively
that
(cf. Fig. 4 in \cite{E:Beam}) by Mironychev and   Babich
\newline \hspace*{7mm} and those (say, cf. Fig. 1 in \cite{1998:Column})
by {Wescott \it et al.} It has  the dendritic type.
\end{figure}

\newpage

\psfig{file=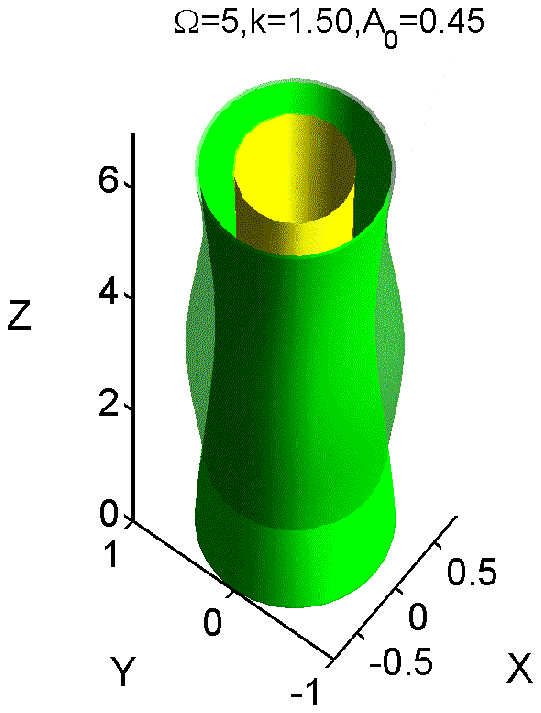,bbllx=0cm,bblly=13.5cm,bburx=12cm,bbury=24cm,rheight=11cm,rwidth=11cm,clip=}
\vspace{6mm}
\begin{figure}[h]
\hspace*{6mm} Fig. 2 \hspace*{1mm} Isosurface plots of the solution
$({\bf e}+{\bf b}) $  in a rotating frame of reference.\newline
\hspace*{7mm}
 Possible column(sprite)-like and helical
electromagnetic field. $\hat{\Omega}=5, \bar{A}_0=0.45, k=1.5$.
\end{figure}

\newpage

\psfig{file=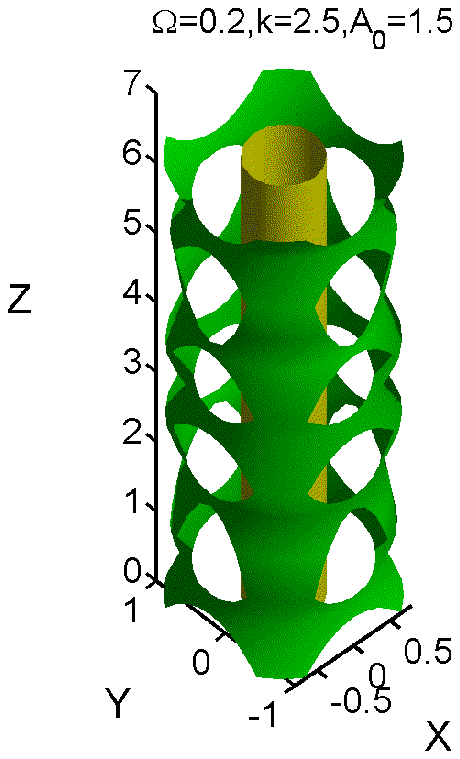,bbllx=0cm,bblly=13.5cm,bburx=12cm,bbury=24cm,rheight=11cm,rwidth=11cm,clip=}
\vspace{6mm}
\begin{figure}[h]
\hspace*{6mm} Fig. 3 \hspace*{1mm} Isosurface plots of the solution
$({\bf e}+{\bf b}) $  in a rotating  frame of reference.
\newline \hspace*{7mm}  Possible sprite- and column-like electromagnetic field.
$\hat{\Omega}=0.2, \bar{A}_0=1.5, k=2.5$.
\newline \hspace*{7mm} (the spherical or green surface is for $|{\cal A}|$)
\end{figure}

\newpage

\psfig{file=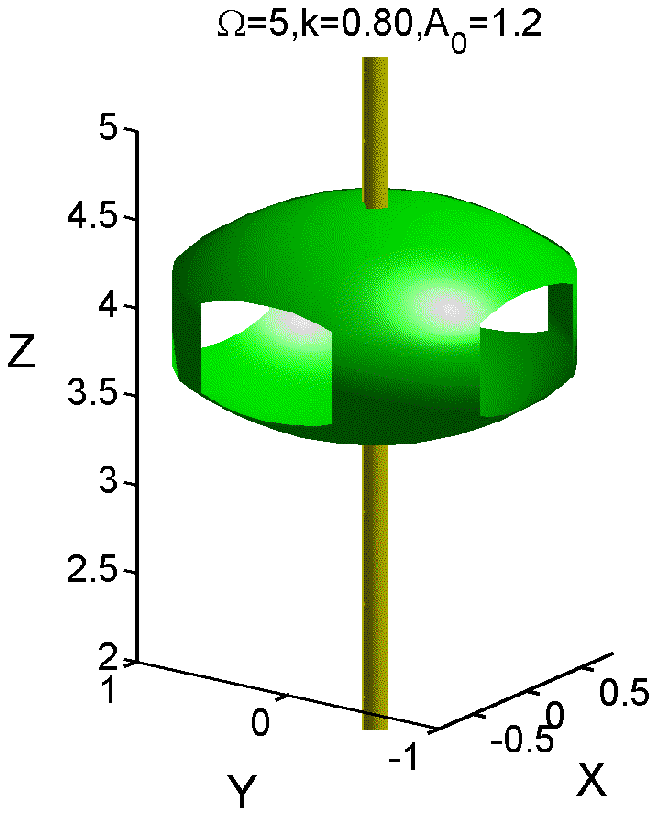,bbllx=0cm,bblly=13.5cm,bburx=12cm,bbury=24cm,rheight=11cm,rwidth=11cm,clip=}
\vspace{6mm}
\begin{figure}[h]
\hspace*{6mm} Fig. 4 \hspace*{1mm} Isosurface plots of the solution
$({\bf e}+{\bf b}) $  in a rotating frame of reference.\newline
\hspace*{7mm}
 Possible sprite and column-like
electromagnetic fields.  $\hat{\Omega}=5.0, \bar{A}_0=1.2, k=0.8$.
\newline \hspace*{7mm} The column-like (yellow) isosurface is
for ${\bf e} ({\bf r})$. This result resembles qualitatively
that \newline \hspace*{7mm}
(cf. Figs. 5 and 6 in \cite{E:Beam}) by Mironychev and   Babich.
\end{figure}

\newpage

\psfig{file=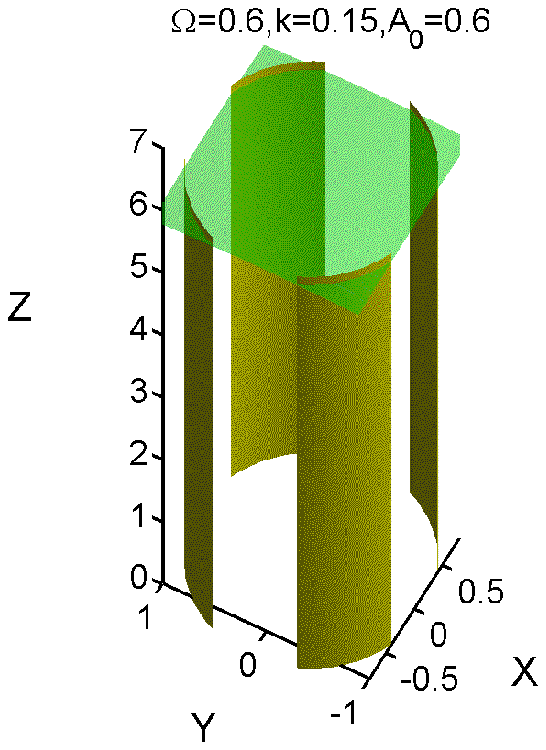,bbllx=0cm,bblly=13.5cm,bburx=12cm,bbury=24cm,rheight=11cm,rwidth=11cm,clip=}
\vspace{6mm}
\begin{figure}[h]
\hspace*{6mm} Fig. 5 \hspace*{1mm} Isosurface plots of the solution
$({\bf e}+{\bf b})$  in a rotating frame  of
 \newline \hspace*{7mm} reference. Possible
sprite-like electromagnetic field  
 $\hat{\Omega}=0.6, \bar{A}_0=0.6,
k=0.15$. \newline \hspace*{7mm} Possible jelly-fish-like pattern
with the downward tendrils-like
feature  appears.  %
\end{figure}

%
\end{document}